\documentclass[preprint,12pt]{elsarticle}

\usepackage{graphicx}
\usepackage{amssymb}
\usepackage{amsmath}
\newtheorem{theorem}{Theorem}

\usepackage{lineno}

\journal{Journal Name}

\begin{document}

\begin{frontmatter}


\title{Effects of Some Operations on Domination Chromatic Number in Graphs}

 \author{Yangyang Zhou}
 \ead{zyy_eecs@pku.edu.cn}
 \author{Dongyang Zhao}
 \address{Peking University, Beijing, China}


\begin{abstract}
For a simple graph $G$, a domination coloring of $G$ is a proper vertex coloring such that every vertex of $G$ dominates at least one color class, and every color class is dominated by at least one vertex. The domination chromatic number, denoted by $\chi_{dd}(G)$, is minimum number of colors among all domination colorings of $G$. In this paper, we discuss the effects of some typical operations on $\chi_{dd}(G)$, such as vertex (edge) removal, vertex (edge) contraction, edge subdivision, and cycle extending. 
\end{abstract}

\begin{keyword}
Domination chromatic number \sep Removal \sep Contraction \sep Edge subdivision \sep Cycle extending

\end{keyword}

\end{frontmatter}


\section{Introduction}
\label{S:1}

Let $G=(V, E)$ be a simple graph. For any vertex $v \in V(G)$, the open neighborhood of $v$ is the set $N(v)=\{u | uv \in E(G)\}$ and the closed neighborhood is the set $N[v]=N(v) \cup \{v\}$. The degree of a vertex $v \in V$, denoted by $deg(v)$, is the cardinality of its open neighborhood. The maximum and minimum degree of a graph $G$ is denoted by $\Delta(G)$ and $\delta(G)$, respectively. For a subset $X \subseteq V$, we denote by $G[X]$ the subgraph of $G$ induced by $X$. We consider only finite, undirected and simple connected graphs in this paper. The reader is referred to the book of Bondy and Murty \cite{Bondy:2008} for any undefined terms.
 
A proper vertex $k$-coloring of a graph $G = (V, E)$ is a mapping $f : V \rightarrow \{1,2,\cdots,k\}$ such that any two adjacent vertices receive different colors. In fact, this problem is equivalent to the problem of partitioning the vertex set of $G$ into $k$ independent sets $\{V_1,V_2,\cdots,V_k\}$ where $V_i = \{x \in V | f(x) = i\}$. The set of all vertices colored with the same color is called a color class. The chromatic number of $G$, denoted by $\chi(G)$, is the minimum number of colors needed in a proper coloring of $G$. 

A dominating set $S$ is a subset of the vertices in $G$ such that every vertex in $G$ either belongs to $S$ or has a neighbor in $S$. The domination number $\gamma(G)$ is the minimum cardinality of a dominating set of $G$. 

Coloring and domination are two important fields in graph theory, and both of them have rich research results. For comprehensive surveys of coloring and domination in graphs, refer \cite{Franklin:1922, Pardalos:1998, Malaguti:2010, Borodin:2013} and \cite{Haynes:199801, Haynes:199802, Ananchuen:2011, Chang:2013}, respectively. Also, relations between the two fields have been discussed in many ways. Chellali and Volkmann \cite{Chellali:2004} showed some relations between the chromatic number and some domination parameters in a graph. Hedetniemi et al. \cite{Hedetniemi:2009} introduced the concept of dominator partition of a graph. Motivated by \cite{Hedetniemi:2009}, Gera et al. \cite{Gera:2006} proposed the dominator coloring as a proper coloring such that every vertex has to dominate at least one color class (possibly its own class) in 2006. Gera researched further on this coloring problem in \cite{Gera:200701, Gera:200702}. Similar to the dominator coloring, Kazemi \cite{Kazemi:2015} studied the concept of a total dominator coloring in 2015, which is a proper coloring such that each vertex of the graph is adjacent to every vertex of some (other) color class. More results on the dominator coloring could be found in \cite{Chellali:2012, Boumediene:2012, ARUMUGAM:2012, Bagan:2017}. Then, in 2015, Boumediene et al. \cite{Boumediene:2015} proposed the dominated coloring as a proper coloring where every color class is dominated by at least one vertex. Based on these studies, we introduce the domination coloring in \cite{Zhou:2019}, in which both the vertices and color classes should satisfy the domination property. 

A domination coloring of $G$ is a proper vertex coloring such that every vertex of $G$ dominates at least one color class (possibly its own class), and every color class is dominated by at least one vertex. The domination chromatic number $\chi_{dd}(G)$ is the minimum number of color classes in a domination coloring of $G$. In \cite{Zhou:2019}, we give basic properties of $\chi_{dd}(G)$, and prove that the $k$-domination coloring problem is NP-Complete for $k \geq 4$.  

With these basic definitions and conclusions, there is another problem that attracts the scholars' attention naturally: What happens to these paramaters, when some operations are implemented on $G$? On this problem, the domination-critical concepts of graphs are proposed to examine the effects of vertex (edge) removal and addition on $\gamma$, more results refer \cite{Sumner:1983, Sumner:1990, Robert:1988, Yamuna:2016, Sumner:2017}. Ghanbari and Alikhani \cite{Ghanbari:2016} study the effects of some operations on the total dominator chromatic number in 2016. In 2018, Alikhani, Ghanbari and Soltani \cite{Alikhani:2018} discuss the total dominator chromatic number of $k$-subdivision of a graph. In this paper, we aim to study the effects on $\chi_{dd}(G)$ when we consider some operations on a graph $G$. In Section 2 and Section 3, we consider the basic removal and contraction operations on the vertex and edge. In Section 4, we examine the effects of edge subdivision. In Section 5, we study the cycle extending operation and its effects on $\chi_{dd}(G)$. In Section 6, we make conclusion and propose the future research.
 
\section{Vertex and edge removal}

This section focuses on the influence of vertex and edge removal on the domination chromatic number. Let $v$ be a vertex and $e$ be an edge of $G$, respectively. The graph $G-v$ is obtained by deleting the vertex $v$ and all edges incident with $v$ in $G$. And $G-e$ is a graph that obtained from $G$ by simply removing the edge $e$. In the following, we discuss the effect on $\chi_{dd}(G)$ when $G$ is changed by the vertex and edge removal. 

\begin{theorem}
\label{th1}
Let $G$ be a connected graph, and $v \in V(G)$ is not a cut vertex, then
$$\chi_{dd}(G)-1 \leq \chi_{dd}(G-v) \leq \chi_{dd}(G)+deg(v)-1.$$
\end{theorem}

Proof. We first prove the left inequality. Given any domination coloring of $G-v$, we consider the domination coloring of $G$ as follows. If we add vertex $v$ and all the corresponding edges to $G-v$, then it suffices to add a new color $i$ to $v$ and all other colors unchanged.  Since $v$ dominates the color class itself and every vertex except $v$ dominates the old color classes, the color class $v$ itself can be dominated by any adjacent vertices of $v$ and all the other color classes be dominated by the original vertices, we obtain a domination coloring for $G$. So,  $\chi_{dd}(G) \leq \chi_{dd}(G-v) + 1$.

Now we prove the right part $\chi_{dd}(G-v) \leq \chi_{dd}(G)+deg(v)-1$. For any domination coloring of $G$, suppose that vertex $v$ have color $i$, then we have two cases:

Case 1. There exsist other vertices with color $i$. If there exsists a color classes which is only dominated by $v$, then we give each vertex of this color class a new color. Obviously, the number of these vertices is at most $deg(v)$. Each of the new colored vertices dominates the color class itself, and each of them as a color class is dominated by another neighbor except $v$, which must exsist, since $v$ is not a cut vertex. So, we get a domination coloring of $G-v$, and $\chi_{dd}(G-v) \leq \chi_{dd}(G)+deg(v)-1$. If there exsist no color class described above, then we can obtain a domination coloring of $G-v$ using the old color of $G$. So, $\chi_{dd}(G-v) \leq \chi_{dd}(G)$.   

Case 2. There exsist no other vertices with color $i$. As the analysis in Case 1, we have $\chi_{dd}(G-v) \leq \chi_{dd}(G)+deg(v)-2$ if there exsist color classes only dominated by $v$, and otherwise, $\chi_{dd}(G-v) \leq \chi_{dd}(G)-1$.

Therefore, $\chi_{dd}(G-v) \leq \chi_{dd}(G)+deg(v)-1$, and the theorem follows.
\hfill $\square$

\begin{theorem}
\label{th2}
Let $G$ be a connected graph, and $e \in E(G)$ is not a cut edge, then
$$\chi_{dd}(G)-1 \leq \chi_{dd}(G-e) \leq \chi_{dd}(G)+2.$$
\end{theorem}

Proof. Let $e=uv \in E$. First, we prove the left inequality. We shall present a domination coloring for $G-e$. If we add the edge $e$ to $G-e$, then there exist two cases. If two vertices $u$ and $v$ have the same color in the domination coloring of $G-e$, then we add a new color i to one of them. Since each vertex use the old color, we obtain a domination coloring for $G$. So we have $\chi_{dd}(G) \leq \chi_{dd}(G-e) + 1$. If two vertices $u$ and $v$ do not have the same color in the domination coloring of $G-e$, then the domination coloring of $G-e$ can also be a domination coloring for $G$. So, $\chi_{dd}(G) \leq \chi_{dd}(G-e)$. Therefore, $\chi_{dd}(G)-1 \leq \chi_{dd}(G-e)$.

Now, we prove $\chi_{dd}(G-e) \leq \chi_{dd}(G)+2$. For a domination coloring of $G$, suppose that $u \in V_i$ and $v \in V_j$, that is $u$ has color $i$ and $v$ has color $j$. Then, we have the following cases:

Case 1. The vertex $u$ does not dominate the color class $V_j$, and the vertex $v$ does not dominate the color class $V_i$. In this case, the domination coloring of $G$ gives a domination coloring of $G-e$. So, $\chi_{dd}(G-e) \leq \chi_{dd}(G)$.

Case 2. The vertex $u$ dominate the color class $V_j$, and the vertex $v$ does not dominate the color class $V_i$. By definition, $u$ adjacents to every vertex in $V_j$, $V_i$ must be dominated by some other vertices. We give a new color $k$ to vertex $v$ in $G-e$. Thus, $u$ dominate the color class formed by $V_j / v$, $v$ dominate the color class itself. Since $e$ is not a cut edge, $G-e$ is connected and $v$ has other neighbors. So there have other vertices dominate the color class $v$ itself. Each other vertex use the old color, so we obtain a domination coloring of $G-e$, and $\chi_{dd}(G-e) \leq \chi_{dd}(G)+1$.

Case 3. The vertex $u$ dominate the color class $V_j$, and the vertex $v$ dominate the color class $V_i$. In this case, we give two new colors $k$ and $l$ to vertices $u$ and $v$, respectively. In the new obtained coloring, both $u$ and $v$ dominate the color class themselves, and there exist other vertices adjacent to $u$ and $v$ by the connectedness of $G-e$, respectively. Since each vertex use the old color, the new coloring is a domination coloring of $G-e$. So, $\chi_{dd}(G-e) \leq \chi_{dd}(G)+2$.

By the above three cases, we have $\chi_{dd}(G-e) \leq \chi_{dd}(G)+2$. So, the result follows.           
\hfill $\square$

\section{Vertex and edge contraction}
 
\begin{figure}[h]
\centering\includegraphics[width=0.5\linewidth]{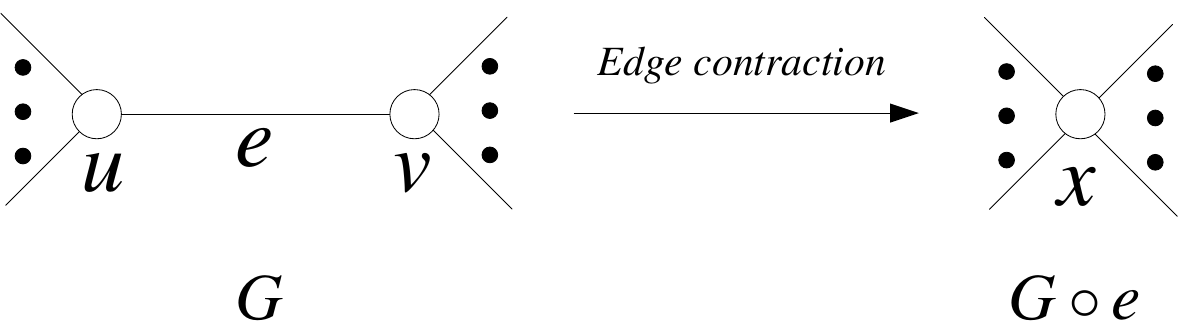}
\caption{The edge contraction operation}
\label{f2}
\end{figure}

For a simple garph $G$, $S \subseteq V(G)$ is a subset of vertices. The contraction of $S$ in $G$, denoted by $G \circ S$, is the graph obtained by replacing all vertices in $S$ with a single new vertex, such that edges insident with the new vertex are those edges that were insident with any vertex in $S$. Specially, for $S=\{u,v\}$, we consider the following two cases:

(1) Vertex $u$ and $v$ are adjacent in $G$, that is $e=uv \in E(G)$. The contraction of an edge $e=uv$ in $G$, denoted by $G \circ e$, is the graph obtained by deleting the edge $e$, and replacing $u$ and $v$ with a new single vertex such that vertices adjacent to the new vertex if they were adjacent to vertex $u$ or $v$, as shown in Figure \ref{f2}, where the black points around the vertex denote that there may be edges insident with the vertex.
 
\begin{figure}[h]
\centering\includegraphics[width=0.66\linewidth]{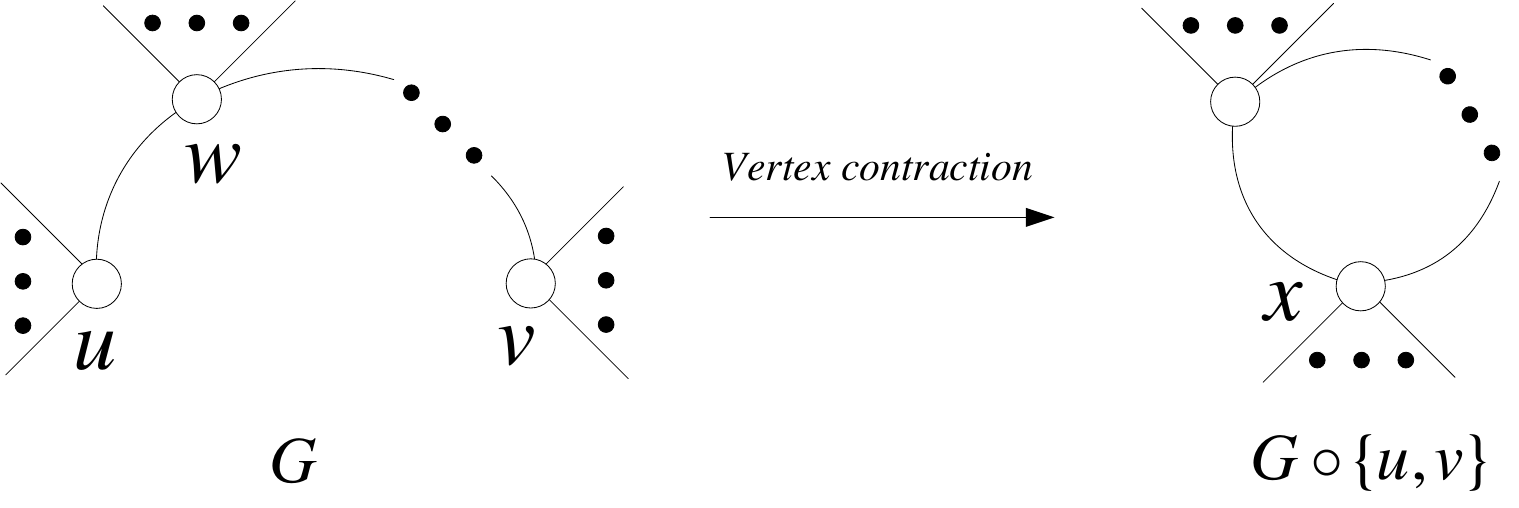}
\caption{The vertex contraction operation}
\label{f3}
\end{figure}

(2) Vertex $u$ and $v$ are not adjacent, $uv \notin E(G)$. We denote this contraction operation by $G \circ \{u,v\}$, as shown in Figure \ref{f3}.  

In this section, we discuss the effect on $\chi_{dd}(G)$ when $G$ is changed by the vertex and edge contraction.

\begin{theorem}
\label{th3}
Let $G$ be a connected graph, $e \in E(G)$, then
$$\chi_{dd}(G)-2 \leq \chi_{dd}(G \circ e) \leq \chi_{dd}(G)+1.$$
\end{theorem}

Proof. Let $e = uv \in E$. We first prove the right inequality. Let $f$ be a $\chi_{dd}(G)$-domination coloring of $G$, $f(u)=i$ and $f(v)=j$. Now we present a domination coloring for $G \circ e$. We denote $x$ the replacement vertex of $u$ and $v$. For $G \circ e$, we give a new color $k$ to vertex $x$ and each other vertices remain the previous coloring in $f$. The obtained coloring is a domination coloring of $G \circ e$, since $x$ dominate the color class itself and this color class must be dominated by any adjacent vertices of $x$, and all other vertices and color classes remain the old domination relationship. So, we have $\chi_{dd}(G \circ e) \leq \chi_{dd}(G)+1$.   

Now, we prove the left inequality. Consider any $\chi_{dd}(G \circ e)$-domination coloring of $G \circ e$. By adding the removed vertex and all the corresponding edges to $G \circ e$, we get the graph $G$. Then remove the old color of the endpoints of $e$, add two new colors $i$ and $j$ to the vertices $u$ and $v$, respectively, and keep the old colors of all other vertices. We get a new coloring for $G$, and it is easy to check this is a domination coloring. So, $\chi_{dd}(G) \leq \chi_{dd}(G \circ e)+2$.

Thus, the result follows.
\hfill $\square$

\begin{theorem}
\label{th4}
For a connected graph $G$, we have
$$\chi_{dd}(G)-2 \leq \chi_{dd}(G \circ \{u,v\}) \leq \chi_{dd}(G)+1.$$
\end{theorem}

Proof. Let $u, v$ are two unadjacent vertices of $G$. We first prove the right inequality. For a $\chi_{dd}(G)$-domination coloring $f$ of $G$, we will present a domination coloring of $G \circ \{u,v\}$ based on $f$. We give a new color $k$ to the new vertex, denoted by $x$, and all the other vertices remain the previous coloring in $f$. Since each vertex other than $x$ satisfies the domination properties, and $x$ dominate the color class itself or the one which is dominated by $u$ or $v$ in $f$, also the color class $\{x\}$ must be dominated by any adjacent vertices of $x$, we obtain a domination coloring of $G \circ \{u,v\}$. So, $\chi_{dd}(G \circ \{u,v\}) \leq \chi_{dd}(G)+1$.

For the left part, we first present a $\chi_{dd}(G \circ \{u,v\})$-domination coloring of $G \circ \{u,v\}$. Then, we add the removed vertex and the corresponding edges to $G \circ \{u,v\}$, in order to get $G$. In $G$, remove the old coloring of $u$ and $v$ and give two new colors to them, and keep other vertices colors unchanged. Like before, we can check this is a domination coloring of $G$. So, we have $\chi_{dd}(G) \leq \chi_{dd}(G \circ \{u,v\})+2$.
    
The result follows.
\hfill $\square$

\section{Edge subdivision}

Edge subdivision is a typical operation in graphs. Let $G$ be a connected graph and $k$ be a positive integer.  The $k$-subdivision of $G$, denoted by $S_k(G)$, is the graph obtained from $G$ by replacing each edge with a path of length $k$. For each edge $e=v_iv_j \in E(G)$, let $P_{v_i,v_j}$ denote the $k$-path that replacing $v_iv_j$, and we call $P_{v_iv_j}$ a superedge, any new vertex of $P_{v_iv_j}$ is an internal vertex. Note that if $k=1$, then $S_1(G)=G$. 
   
In this section, we investigate the domination chromatic number of the $k$-subdivision of graph $G$, and the relation between $\chi_{dd}(G)$ and $\chi_{dd}(S_k(G))$.

\begin{theorem}
\label{th5}
Let $G$ be a connected graph with $m$ edges, and $k \geq 2$, then
$$\chi_{dd}(P_{k+1}) \leq \chi_{dd}(S_k(G)) \leq (m-1)\chi_{dd}(P_k) + \chi_{dd}(P_{k+1}).$$
\end{theorem}

\begin{figure}[h]
\centering\includegraphics[width=0.88\linewidth]{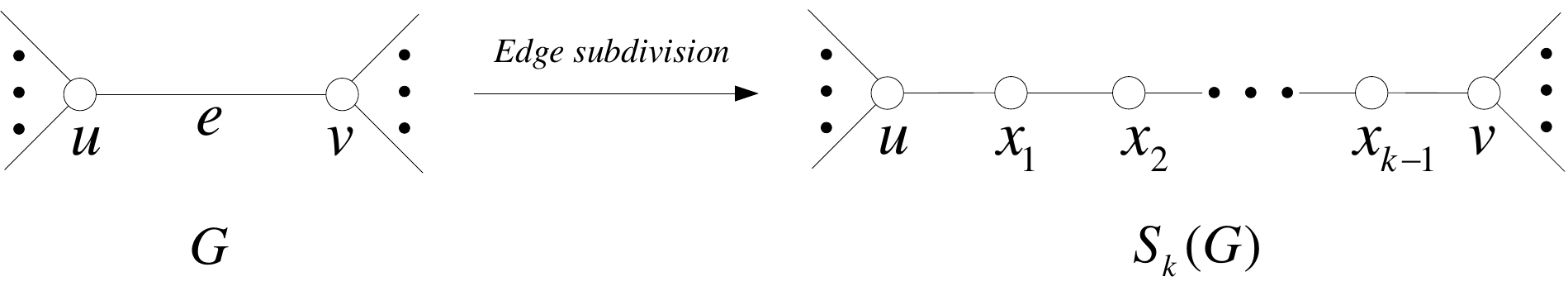}
\caption{The edge subdivision operation}
\label{f4}
\end{figure}

Proof. We first prove the left inequality. Let $e=uv$ be an arbitrary edge of $G$, and $P_{uv}=ux_1x_2 \cdots x_{k-1}v$ be the replacement of $uv$ in $S_k(G)$, as shown in Figure \ref{f4}. We shall get a domination coloring of $P_{k+1}$ base on a domination coloring of $S_k(G)$. Let $f$ a $\chi_{dd}$-domination coloring of $S_k(G)$. We consider the restriction of $f$ to the induced subgraph $P_{uv}$ and denote it by $f'$. In $f$, both of the endpoints $u$ and $v$ dominate either the color class itself or the color class formed by some adjacent vertices of them, respectively. Thus, they also dominate the color class itself or the color class formed by some adjacent vertices of them in $f'$. With the internal vertices and their colorings unchanged, we conclude that each color class of $f'$ must be dominated by the vertex itself or the same vertex as in $f$. So, $f'$ is a domination coloring of $P_{uv}$, and we have $\chi_{dd}(P_{k+1}) \leq |f'| \leq |f| = \chi_{dd}(S_k(G))$.

Now, we prove the right inequality. Let $G$ be a connected graph with $m$ edges. For any vertex $u \in V(G)$, let $N_G(u)=\{v_1,v_2, \cdots ,v_p\}$. The edges $uv_i$ are replaced with the superedges $P_{uv_i}$$(i=1,2,\cdots,p)$, respectively. We can easily color the vertices of superedge $P_{uv_1}$ with $\chi_{dd}(P_{k+1})$ colors as a domination coloring of $P_{k+1}$. Then, we color the vertices of $P_{uv_i}$, $i=1,2,\cdots,p$, such that vertex $u$ has the unique color in $P_{uv_i}(i=2,\cdots,p)$ as the color it has in $P_{uv_1}$, and the following formula is satisfied:
$$\{f(P_{uv_i}) \cap f(P_{uv_j})\} \backslash \{f(u)\} = \emptyset,$$
where $i, j = 1,2,\cdots,p$ and $i \neq j$, $f(P_{uv_i})$ is the set of colors of vertices in $P_{uv_i}$. 
Thus, we obtain a domination coloring of $P_{uv_i}$$(i=1,2,\cdots,p)$ with at most $(p-1)\chi_{dd}(P_k) + \chi_{dd}(P_{k+1})$ colors. Next, we consider the superedges which are uncolored and are replacements of edges incident with vertices $v_i$, $i=1,2,\cdots,p$. Continue the above process until all vertices of $S_k(G)$ are colored. Note that some vertices will be recolored when they lie on some cycles. The coloring obtained finally is a domination coloring of $S_k(G)$, which use at most $(m-1)\chi_{dd}(P_k) + \chi_{dd}(P_{k+1})$ colors. So, we get the right part and the theorem follows. 
\hfill $\square$

 

\section{Cycle extending}

\begin{figure}[h]
\centering\includegraphics[width=0.7\linewidth]{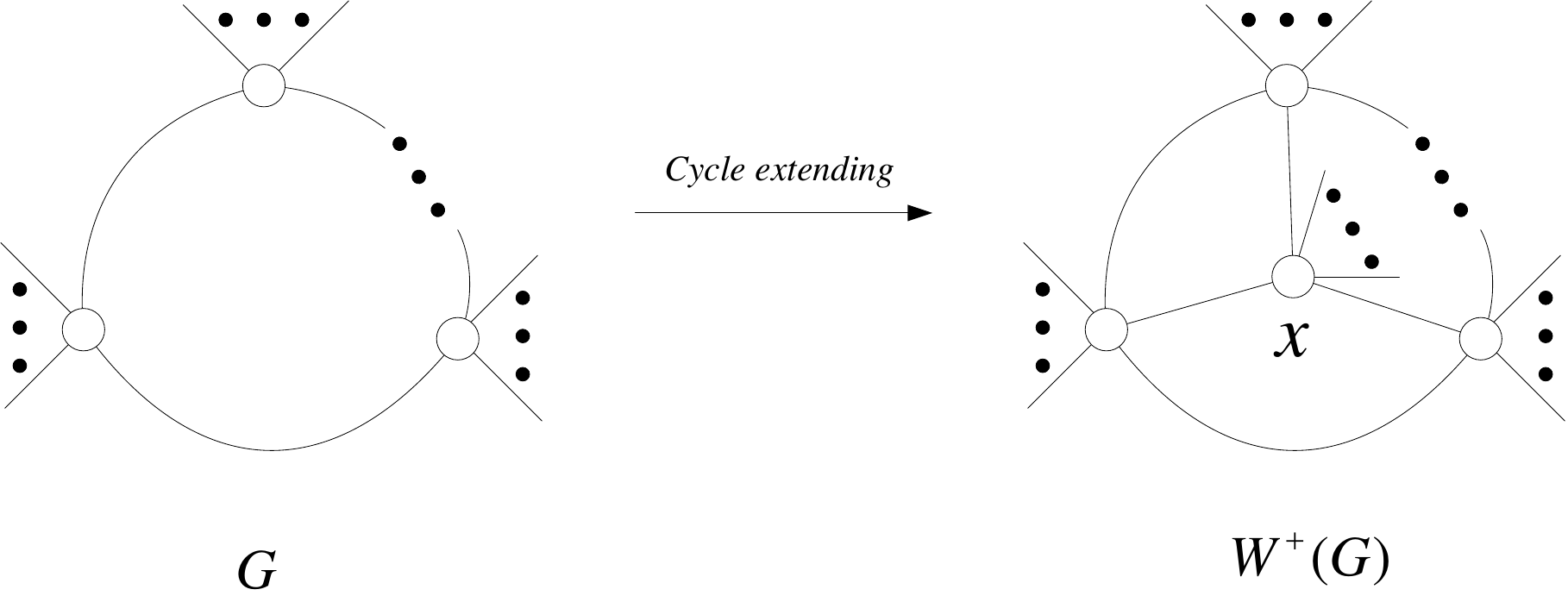}
\caption{The cycle extending operation}
\label{f1}
\end{figure}

In thia section, we focus on the cycle extending operation in graphs, and study its influence on the domination chromatic number. For a connected graph $G$, $C$ is a cycle of length $l(\geq 3)$ in $G$. The cycle extending based on $C$ is adding a new vertex $x$ and connecting $x$ to every vertex on $C$. We denote the obtained graph by $W^+_G(G)$, which is shorted for $W^+(G)$ without confusion. The operation is shown in Figure \ref{f1}, where the black points around the vertices denote that there may be edges, and the black points between edges denote that the length of cycle $C$ is at least 3.

\begin{theorem}
\label{th6}
Let $G$ be a connected graph and $C$ be any cycle of length $l$ in $G$, then
$$\chi_{dd}(G)-l \leq \chi_{dd}(W^+(G)) \leq \chi_{dd}(G) + 1.$$
\end{theorem}

Proof. We first consider the right part of the inequality. Let $f$ be a domination coloring of $G$. We can get a domination coloring of $W^+(G)$ based on $f$ by adding a new color to the center vertex $x$ and remaining other colors of vertices unchanged. Thus, $\chi_{dd}(W^+(G)) \leq \chi_{dd}(G) + 1$.

Now, we prove the left part. Let $f^*$ be a domination coloring of $W^+(G)$. We aim to get a domination coloring of graph $G$ based on $f^*$. There exsist two cases for the color of the center vertex $x$ in $f^*$.

Case 1. $x$ has the unique color in $f^*$, that is, $x$ itself is a color class. In this case, we need to consider the vertices which only dominate the color class formed by $x$ and the color classes which are only dominated by $x$, and all these elements are vertices on cycle $C$. So, just adding at most $l$ new coloring to these vertices, we can obtain a domination coloring of $G$. The inequality follows.

Case 2. There exsist some other vertices having the same color as $x$. We denote the color class containing $x$ by $V_i$. Since the domination properties of $V_i$ and all other vertices are not affected by the deletion of $x$, we need to consider the color classes that are dominated by $x$ only. Also, all these color classes must be formed by vertices on cycle $C$. Thus, we get a domination coloring of $G$ by adding at most $l$ colors like Case 1. The left inequality is obtained.

So, the theorem follows.
\hfill $\square$



\section{Conclusion and Further Research}

In this paper, we study the effects on the domination chromatic number $\chi_{dd}(G)$ of a graph $G$ when $G$ is modified by some operations related to the vertices, edges and cycles of $G$. Specificly, several general bounds and properties of the new created graphs are given with respect to $\chi_{dd}(G)$.

For the future research, we plan to examine the effects of some complex operations on the domination chromatic number, such as the edge filping operation, the complementary operation and some binary operations. Also, we are interest in the operations which do not affect the original domination chromatic number.

\noindent\textbf{Acknowledgements}

The authors acknowledge support from Peking University. The authors also would like to thank referees for their useful suggestions. 





\bibliographystyle{model1-num-names}
\bibliography{sample}







\end{document}